\documentclass[aps,prl,twocolumn,superscriptaddress,longbibliography]{revtex4-2}
\usepackage{graphicx}
\usepackage{amssymb,amsmath}
\usepackage{bm}
\usepackage{dcolumn}
\usepackage{float}
\usepackage[OT1]{fontenc} 
\usepackage{url}
\usepackage{mathrsfs}
\usepackage{slashed,comment}
\usepackage{color}
\usepackage{verbatim}
\usepackage{enumitem}
\usepackage{soul,physics}
\usepackage[driverfallback=dvipdfm]{hyperref}
\hypersetup{pdfpagemode=FullScreen,colorlinks=true,breaklinks,urlcolor=blue,linkcolor=blue,citecolor=blue}

\usepackage{subfigure}
\usepackage{amssymb}
\usepackage{bm}
\usepackage{graphicx}
\usepackage{amsmath,amssymb}

\begin{document}
 
  \title{Universal Relations in Long-range Quantum Spin Chains }

  \author{Ning Sun}
  \affiliation{Department of Physics, Fudan University, Shanghai, 200438, China}

  \author{Lei Feng}
  \thanks{leifeng@fudan.edu.cn}
  \affiliation{Department of Physics, Fudan University, Shanghai, 200438, China}
  \affiliation{State Key Laboratory of Surface Physics, Fudan University, Shanghai, 200438, China}
  \affiliation{Institute for Nanoelectronic devices and Quantum computing, Fudan University, Shanghai, 200438, China}
  \affiliation{Shanghai Key Laboratory of Metasurfaces for Light Manipulation, Shanghai, 200433, China}
  \affiliation{Hefei National Laboratory, Hefei 230088, China}

  \author{Pengfei Zhang}
  \thanks{PengfeiZhang.physics@gmail.com}
  \affiliation{Department of Physics, Fudan University, Shanghai, 200438, China}
  \affiliation{State Key Laboratory of Surface Physics, Fudan University, Shanghai, 200438, China}
  \affiliation{Hefei National Laboratory, Hefei 230088, China}

  \date{\today}

  \begin{abstract}
  Understanding the emergence of novel collective behaviors in strongly interacting systems lies at the heart of quantum many-body physics. Valuable insight comes from examining how few-body correlations manifest in many-body systems, embodying the ``from few to many'' philosophy. An intriguing example is the set of universal relations in ultracold atomic gases, which connect a wide range of observables to a single quantity known as the contact. In this Letter, we demonstrate that universal relations manifest in a distinct class of quantum many-body systems, long-range quantum spin chains, which belong to a completely new universality class. Using effective field theory and the operator product expansion, we establish connections between the asymptotic behavior of equal-time spin correlation functions, the dynamical structure factor, and the contact density. The theoretical predictions for equal-time correlators are explicitly verified through numerical simulations based on matrix product states. Our results could be readily tested in state-of-the-art trapped-ion systems.
  \end{abstract}
    
  \maketitle

  \emph{ \color{blue}Introduction.--} Strongly interacting quantum systems pose significant challenges to understanding collective phenomena, since conventional perturbative theory breaks down in the absence of a small expansion parameter. Nevertheless, progress can be made when the system exhibits a separation of length scales, with the average separation between particles much larger than the interaction range. In this regime, probing the system at short distances or high momenta reveals that the dominant contributions originate from universal few-body physics \cite{Hammer:2005bp,zhai2021ultracold}. Following this idea, a set of elegant universal relations has been proposed for two-component Fermi gases, connecting many observables, such as the momentum distribution and the radio-frequency spectrum, to a central quantity called the contact \cite{2008AnPhy.323.2952T,2008AnPhy.323.2971T,2008AnPhy.323.2987T,PhysRevLett.100.205301,PhysRevLett.99.190407,PhysRevLett.99.170404,2009EPJB...68..401W,PhysRevA.79.053640,PhysRevA.79.023601,PhysRevA.81.063634,PhysRevLett.104.223004}. These relations have already been experimentally validated in ultracold atomic gases \cite{PhysRevLett.95.020404,PhysRevLett.104.235301,PhysRevLett.105.070402,PhysRevLett.106.170402,PhysRevLett.109.220402,PhysRevLett.110.055305,PhysRevLett.121.093402,PhysRevLett.122.203401}. Subsequent developments have prompted extensive theoretical analyses, including systems exhibiting high partial-wave resonances \cite{PhysRevLett.115.135303,PhysRevLett.115.135304,PhysRevLett.116.045301,PhysRevA.95.043609}, universal three-body correlations \cite{PhysRevLett.106.153005,PhysRevA.83.063614,PhysRevLett.112.110402,PhysRevA.95.033611,PhysRevA.96.030702}, or spin-orbit couplings \cite{PhysRevLett.120.060408,PhysRevA.97.053602,PhysRevA.97.040701,PhysRevA.101.043616,PhysRevA.102.043321,PhysRevA.101.063619}.

  \begin{figure}[t]
    \centering
    \includegraphics[width=0.85\linewidth]{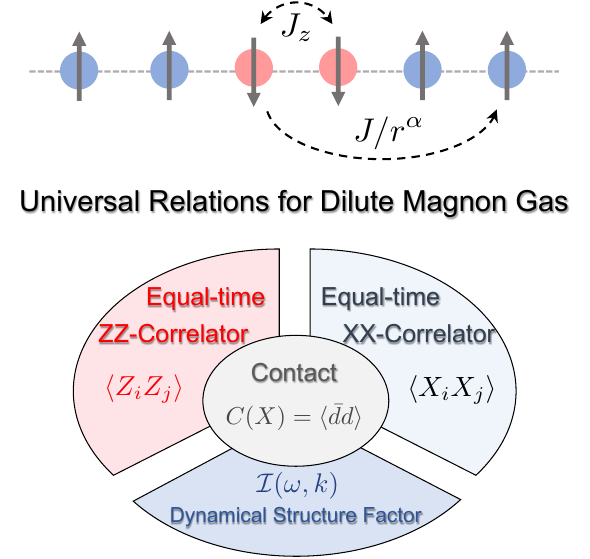}
    \caption{We present a schematic of our main results. We consider a one-dimensional long-range quantum spin chain with spin-rotation symmetry along the $z$-direction, focusing on $\alpha\in(3/2,2)$, where the system is strongly interacting near the two-magnon resonance. In the dilute magnon regime, we derive universal relations that connect equal-time spin correlation functions and the dynamical structure factor to the contact density. 
    Although the system lacks full spin-rotation symmetry, the spin correlators of different type share the same contact constant $C$. }
    \label{fig:schemticas}
  \end{figure}
  
  While these studies focus on non-relativistic particles with short-range interactions, recent advances in quantum science and technology have highlighted novel quantum simulation platforms where the effective degrees of freedom are quantum spins with long-range couplings. Notable examples include trapped-ion systems \cite{wu2023research,2019ApPRv...6b1314B,Foss-Feig:2024blk,2021NatRM...6..892B,10164086,RevModPhys.93.025001,RevModPhys.93.025001}, Rydberg atom arrays \cite{Evered:2023aa,Ma:2023aa,Bluvstein:2024aa,Bekenstein:2020aa,Bluvstein:2021aa,Ebadi:2022aa,Bluvstein:2022aa,Lis:2023aa,Manetsch:2024aa,Tao:2024aa,Cao:2024aa}, and solid-state NMR systems \cite{JONES2001325,RevModPhys.76.1037,Lu2015NMRQI,Cory2000NMRBQ,Laflamme2002IntroductionTN}. When the system exhibits spin-rotation symmetry along the $z$-direction, the number of magnon excitations identified with down spins is conserved. The long-range couplings give rise to a generic low-energy magnon dispersion, $\epsilon_k = u |k|^z$, with a tunable dynamical exponent $z\in(0,2]$ \cite{LEPORI201635,PhysRevB.94.125121,PhysRevA.93.043605,2025arXiv250220759S}. This enables the study of quantum phenomena with no analog in systems of non-relativistic particles, significantly extending our understanding of universality. For instance, Refs.~\cite{2025arXiv250220759S,2025arXiv250704993S} have unveiled universal few-body states that lie beyond the traditional paradigm. However, their implications for many-body physics remain unexplored.

  In this Letter, we establish a set of universal relations for long-range quantum spin systems in the dilute-magnon regime. We focus on physical observables directly relevant to experiments on quantum simulation platforms, including equal-time spin correlators and the dynamical structure factor. By applying the operator product expansion (OPE) within the effective field theory framework, we relate all these observables to the contact density, which quantifies the probability of two magnons approaching each other. We provide direct numerical verification using matrix product state (MPS) simulations on systems of moderate size for universal relations of equal-time correlators. Our results demonstrate how few-body correlations give rise to many-body phenomena in systems with a generic dynamical exponent. We expect that our theoretical predictions could be experimentally tested in trapped-ion systems.

  \emph{ \color{blue}Model \& Effective Field Theory.--} We consider quantum spin chains with long-range couplings in the $x$- and $y$-direction \cite{RevModPhys.93.025001}, and short-range couplings in the $z$-direction \cite{Schuckert:2023pej}. The microscopic Hamiltonian is given by
  \begin{equation}\label{eqn:microscopic}
  H=-\sum_{i,r>0}\bigg[\frac{J}{r^\alpha}\left(X_i X_{i+r}+Y_iY_{i+r}\right)+f(r)Z_i Z_{i+r}\bigg].
  \end{equation}
  Here, we allow a generic short-range coupling $f(r)$ that decays exponentially at large $r$. For conciseness, we set the lattice constant $a=1$. The spin model has an equivalent description in terms of magnons. We choose the fully polarized state $|0\rangle \equiv \otimes_j \ket{\uparrow}_j$ as the vacuum, with a down spin representing a magnon excitation. This identification maps the spin lowering operator $S^-_j = (X_j - iY_j)/2$ to the magnon creation operator $\psi^\dagger_j$, and the spin operator $Z_j$ to the magnon density $n_j=\psi_j^\dag\psi_j$ as $Z_j=1 - 2n_j$. Therefore, the coupling along the $z$-direction becomes a short-range interaction between magnons \footnote{The constraint prohibiting two magnons from occupying the same site can also be implemented by introducing an infinite on-site repulsion.}. Taking the continuum limit gives rise to an effective field theory that governs the universal low-energy properties near the two-body resonance \cite{2025arXiv250220759S}:
  \begin{equation}\label{eq:fieldtheory}
  L=\sum_k \bar{\psi}_k(i\partial_t-\epsilon_k )\psi_k-\frac{1}{2}\int dx~\Big(\bar{\psi}\bar{\psi}d+\bar{d}{\psi}{\psi}-\frac{\bar{d}d}{g}\Big).
  \end{equation}
  The model is defined with a momentum cutoff $\Lambda \sim O(1/a) = O(1)$. $d$ denotes the dimer field, which mediates the contact interaction between magnons \cite{Bedaque:1998kg, Bedaque:1998km}. Higher-order terms involving additional space-time gradients have been omitted. The dispersion of magnons at small momentum $|k|\ll 1$ reads
  \begin{equation}
  \epsilon_k=4\sum_{r=1}^\infty\left[-\frac{J}{r^\alpha} \cos(kr)+f(r)\right] \approx \epsilon_0+ u |k|^z,
  \end{equation}
  where $z = \min\{\alpha - 1, 2\}$ for $\alpha > 1$, and an explicit expression for $u$ is derived in \cite{2025arXiv250220759S}. Since the magnon number is conserved, we omit $\epsilon_0$ and take $\epsilon_k = u |k|^z$ in the following discussion. The validity of the effective field theory has been explicitly demonstrated in the two-body and three-body sectors for $f(r) = J_z \delta_{r1}$ \cite{2025arXiv250220759S}. We emphasize that for many-magnon states, the low-energy approximation requires a large average separation between magnons, or equivalently, a low magnon density $\bar{n}$. Quantum states that satisfy this condition are referred to as a dilute magnon gas.
  
  \begin{figure}[t]
    \centering
    \includegraphics[width=0.92\linewidth]{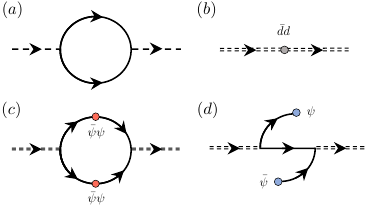}
    \caption{ Feynman diagrams for (a) the self-energy of the dimer field $d$, (b) the matrix element of $\bar{d}d(x)$, (c) the matrix element of $\bar{\psi}\psi(x_1)\bar{\psi}\psi(x_2)$, and (d) the matrix element of $\bar{\psi}(x_1)\psi(x_2)$. In all diagrams, solid and dashed lines denote the bare propagators of $\psi$ and $d$, respectively. Double dashed lines denote the dressed propagator of the dimer field $d$.  }
    \label{fig:diagrams}
  \end{figure}

  We primarily focus on $\alpha\in(3/2,2)$, where the two-magnon problem is renormalizable with the renormalization relation 
  \begin{equation}
  \frac{1}{2g}=\frac{1}{P}-\frac{\Lambda^{1-z}}{4\pi u(1-z)}
  \end{equation}
  with low-energy scattering parameter $P$. The scattering $T$-matrix $T(E,k)$ for a pair of magnons with total energy $E$ and total momentum $k$ is equivalent to the dressed propagator of the dimer field. The self-energy of the dimer field is shown in FIG. \ref{fig:diagrams}(a), which leads to $T(E,k)^{-1}={P}^{-1}-\Sigma_r(E,k)$ with
\begin{equation}
   \Sigma_r(E,k)=\frac{1}{2}\int \frac{dq}{2\pi}\left[\frac{1}{E_+-\epsilon_{\frac k2+q}-\epsilon_{\frac k2-q}}+\frac{1}{2u|q|^z}\right],
  \end{equation}
  where we introduced $E_+=E+i0^+$ for conciseness. Diagrammatically, the dressed propagator of the dimer field is represented by double dashed lines.
  
  \emph{ \color{blue}Equal-time Correlators.--} We are primarily interested in physical observables that are experimentally relevant for quantum simulation platforms. As a first example, we consider the equal-time $ZZ$-correlator $\langle Z_i Z_j \rangle$, where the expectation value is evaluated in a low-energy state within the dilute magnon limit, either in or out of thermal equilibrium. We are interested in extracting its universal behavior, which originates from few-body physics, in the regime where $|i-j|$ is much larger than the lattice constant but much smaller than the average distance between magnons $\bar{n}^{-1}$ \footnote{Moreover, $|i-j|$ should also be much smaller than other many-body length scales, such as $(T/u)^{-1/z}$ in thermal equilibrium at temperature $T$. Similar requirements are kept implicit in later discussions.}. In the effective field theory, this amounts to analyzing the short-distance behavior of $\bar{\psi}\psi(x_1)\bar{\psi}\psi(x_2)$ using the OPE \cite{Wilson:1969zs}, which seeks an operator relation
  \begin{equation}
  \bar{\psi}\psi(x_1)\bar{\psi}\psi(x_2)= \sum_if_{O_i}(x)O_i(X).
  \end{equation}
  Here, we introduce $x=x_1-x_2$ and $X=(x_1+x_2)/2$. ${O_i(X)}$ denotes a set of local operators, and the expansion function $f_{O_i}(x)$ is known as the Wilson coefficient. In particular, we are interested in finding the local operator that dominates at short distances, with a Wilson coefficient that is non-analytic near $x = 0$.

  To calculate the OPE, we first examine the matrix element of $\bar{\psi}\psi(x_1)\bar{\psi}\psi(x_2)$ between an incoming dimer with energy $E$ and momentum $k$ and an outgoing dimer with energy $E'$ and momentum $k'$. Diagrammatically, the insertion of $\bar{\psi}\psi$ induces the scattering of magnons, represented by the red dots in Fig.~\ref{fig:diagrams}(c). Leaving the detailed calculations to the supplementary material \cite{SM}, the leading-order contribution at short distances reads
  \begin{equation}\label{eq:expansion_nn}
  \langle\bar{\psi}\psi(x_1)\bar{\psi}\psi(x_2)\rangle \approx \frac{\Gamma(1-z)^2\sin^2(\frac{\pi z}{2})}{4u^2\pi^2}\frac{e^{i (k-k')X}}{|x|^{2-2z}}.
  \end{equation}
  Here, operators are inserted at $t=0$. $\Gamma(x)$ denotes the Euler gamma function, and the external propagators are not included. Since $2-2z>0$ for $\alpha \in (3/2,2)$, the result appears to diverge at small $|x|$. This divergence signals the breakdown of the effective field theory at distances of a few lattice constants and should be cut off for $|x|\Lambda \lesssim 1$. Next, we identify the corresponding local operator $O(X)$ by matching the $X$-dependence for the matrix element. In particular, the simple phase dependence for arbitrary $k$ and $k'$ motivates the study of the operator $\bar{d}d(X)$, which represents a local scattering potential for dimers. The matrix element is represented by the diagram shown in FIG. \ref{fig:diagrams}(b) and the result is simply given by $\langle\bar{d}d(X)\rangle=e^{i (k-k')X}$. This allows us to determine the Wilson coefficient
  \begin{equation}
  f_{\bar{d}d}(x)=\frac{\Gamma(1-z)^2\sin^2(\frac{\pi z}{2})}{4u^2\pi^2}\frac{1}{|x|^{2-2z}}.
  \end{equation}
  We emphasize that since the OPE is a relation between operators, it holds as long as the effective field theory description remains valid, both in and out of thermal equilibrium. Therefore, by relating the magnon density to spin operators and introducing $c(X)=\langle \bar{d}d(X)\rangle_\rho$ for a generic density matrix $\rho$ of dilute magnon gases, the corresponding equal-time $ZZ$-correlator satisfies the universal relation in the regime $1 \ll |i-j| \ll \overline{n}^{-1}$:
  \begin{equation}\label{eq:resZZ}
  \langle Z_iZ_j\rangle_\rho\approx4\langle n_in_j\rangle_\rho\approx \frac{\Gamma(1-z)^2\sin^2(\frac{\pi z}{2})}{u^2\pi^2} \frac{c(X)}{|x|^{2-2z}},
  \end{equation}
  with $X=(i+j)/2$. Here, $c(X)$ is referred to as the contact density in the context of ultracold atomic gases \cite{PhysRevLett.95.020404,PhysRevLett.104.235301,PhysRevLett.105.070402,PhysRevLett.106.170402,PhysRevLett.109.220402,PhysRevLett.110.055305,PhysRevLett.121.093402,PhysRevLett.122.203401}, and it depends on the state of the system. As we will see, the same quantity also governs the asymptotic behavior of other correlators.

  The universal behavior of $XX$-correlator can be established using the same approach. In terms of the effective field theory, the $XX$-correlator $\langle X_iX_j\rangle$ corresponds to $\langle \bar{\psi}(x_1)\psi(x_2) + \psi(x_1)\bar{\psi}(x_2)\rangle$. Following similar OPE analysis, the matrix element between dimer states is shown in FIG.~\ref{fig:diagrams}(d). Leaving the details to the supplementary material \cite{SM}, the dominant non-analytic contribution yields
  \begin{equation}
  \langle \bar{\psi}(x_1)\psi(x_2)\rangle\approx \frac{\Gamma(1-2z)\sin({\pi z})}{4u^2\pi}\frac{e^{i (k-k')X}}{|x|^{1-2z}}.
  \end{equation}
  Unlike the result \eqref{eq:expansion_nn} for magnon density operators, this non-analytic term vanishes as $|x|\rightarrow 0$. Transforming back to the microscopic spin model and matching the matrix elements of $\bar{d}d(X)$, we obtain the universal relation for the $XX$-correlator for as
  \begin{equation}\label{eq:resXX}
  \langle X_iX_j\rangle_\rho\approx2\langle S^+_iS^-_j\rangle_\rho\approx \frac{\Gamma(1-2z)\sin({\pi z})}{2u^2\pi} \frac{c(X)}{|x|^{1-2z}}.
  \end{equation}
  Due to spin-rotation symmetry along the $z$-direction, the same expression holds for $\langle Y_i Y_j \rangle_\rho$. Equations \eqref{eq:resZZ} and \eqref{eq:resXX} share the same coefficient $c(X)$, providing a nontrivial consistency condition for the universal relations.
  
  \begin{figure}[t]
    \centering
    \includegraphics[width=0.99\linewidth]{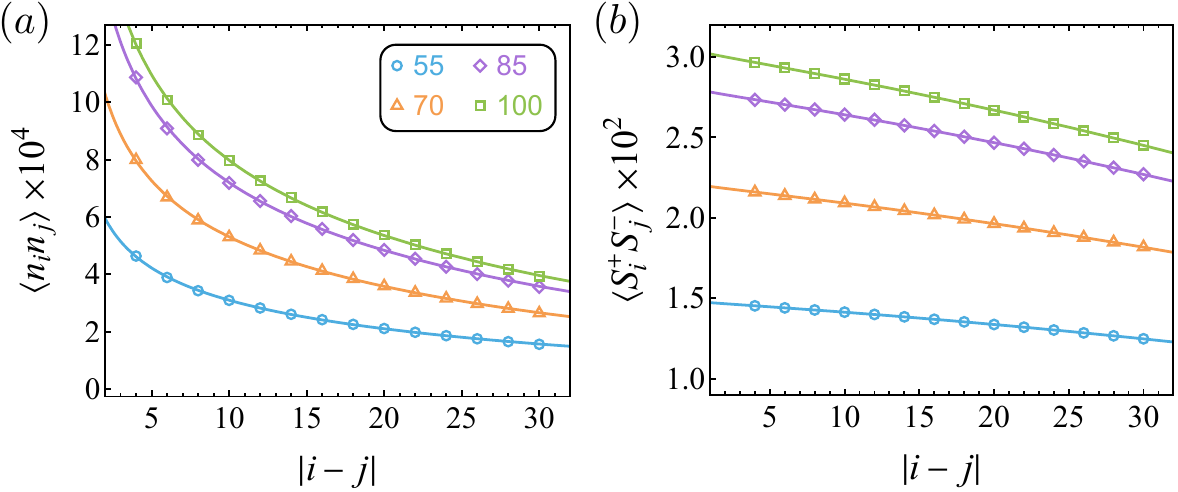}
    \caption{ Numerical results for equal-time correlators with $\alpha=1.9$ and $J_z/J=1.9$ in a system of size $L=200$ with open boundary conditions, evaluated in the ground state of the three-magnon sector ($N=3$). The correlators are shown as functions of $|i-j|$ for different values of $X=(i+j)/2$. Data points represent numerical results, while solid lines correspond to fits over $|i-j| \in [4,30]$, as elaborated in the main text.}
    \label{fig:numfit}
  \end{figure}

  \begin{figure}[t]
    \centering
    \includegraphics[width=0.95\linewidth]{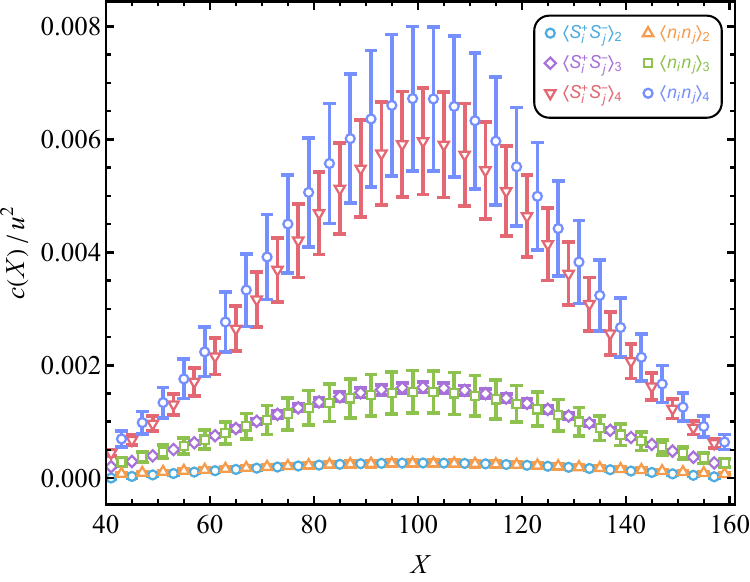}
    \caption{ Results for the contact density $c(X)$, extracted from fits to correlators in the ground state of the system with $N=2,3,4$ magnons. We set $\alpha=1.9$, $J_z/J=1.9$, and consider a system of size $L=200$ with open boundary conditions. Error bars are estimated by varying the fitting range $|i-j| \in [L_s,30]$ with $L_s \in \{2,4,6,8\}$.  }
    \label{fig:res}
  \end{figure}

  \emph{\color{blue}Numerics.--} 
  After establishing the universal relations for equal-time correlators, we numerically validate these predictions in the ground state of the few-magnon sectors using the MPS algorithm with spin-rotation symmetry, an approach that remains efficient for extracting ground-state properties even in long-range coupling systems. The simulations are performed with the \texttt{ITensors.jl} package \cite{10.21468/SciPostPhysCodeb.4}. We adopt the simplest form of couplings along the $z$-direction, $f(r) = J_z \delta_{r1}$, and tune $J_z$ to induce the two-magnon resonance.

  Our simulations are performed on a system of moderate size, $L=200$, consistent with current trapped-ion experiments \cite{wu2023research,2019ApPRv...6b1314B,Foss-Feig:2024blk,2021NatRM...6..892B,10164086,RevModPhys.93.025001}. We set $\alpha=1.9$ and $J_z/J=1.9$ \cite{2025arXiv250220759S}, values close to the two-body resonance $J_z^*/J\approx1.84$. To fulfill both the low-energy and dilute-magnon conditions, we focus on the ground state with magnon numbers $N=2,3,4$. After obtaining the ground state in the corresponding sector, we compute the correlation function matrices $\langle n_i n_j \rangle_N$ and $\langle S_i^+ S_j^- \rangle_N$. The results for $N=3$ are shown in FIG. \ref{fig:numfit} as a function of $|i-j|$ for several values of $X = (i+j)/2$. To test the universal relations, we fit the results of $\langle n_i n_j \rangle$ using $a_f + b_f |i-j|^{2z-2}$. The results shown in FIG. \ref{fig:numfit}(a) exhibit good agreement over the fitting range $|i-j| \in [4,30]$. For $\langle S_i^+ S_j^- \rangle$, the non-analytic term vanishes as $|i-j| \rightarrow 0$, so it is necessary to retain the subleading contribution \cite{SM}. Accordingly, we fit the results using $ \tilde{a}_f + \tilde{b}_f |i-j|^{2z-1} (1 + \tilde{e}_f |i-j|^z)$, which matches the numerical data accurately. Similar results are obtained for other magnon numbers. Eq. \eqref{eq:resZZ} and \eqref{eq:resXX} then relate both $b_f$ and $\tilde{b}_f$ to the contact density $c(X)/u^2$. We plot the results obtained from both correlators in FIG. \ref{fig:res} as a function of $X$ for different magnon numbers $N = 2, 3, 4$. The results clearly demonstrate consistency within the error bars, which are determined by varying the fitting range $|i-j| \in [L_s,30]$ with $L_s \in \{2,4,6,8\}$.

\begin{figure}[t]
    \centering
    \includegraphics[width=0.8\linewidth]{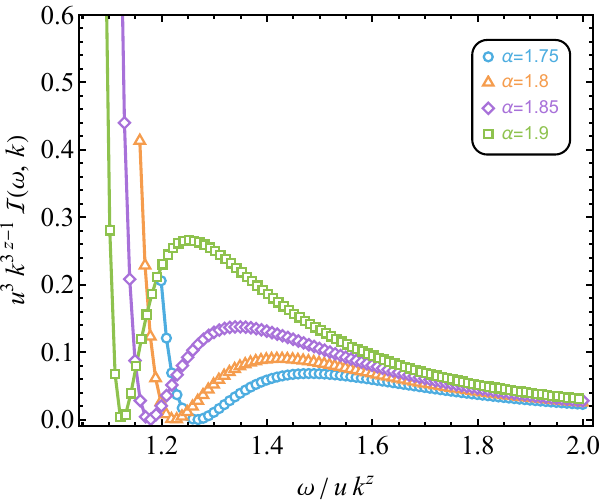}
    \caption{ We plot the universal behavior of the dynamical structure factor as a function of $\omega / (u k^z)$ for various $\alpha$, which is non-vanishing only for $\omega/uk^z>1/2^{z-1}$. The integral expression are provided in the supplementary material \cite{SM}.   }
    \label{fig:DSF}
  \end{figure}

  \emph{\color{blue}Dynamical Structure Factor.--} Finally, we consider another widely used experimental probe in quantum simulation platforms: spectroscopy, which reveals the fundamental properties of interacting many-body systems. In particular, we focus on the dynamical structure factor, which characterizes the system’s response to a time-dependent magnetic field $\delta H = B_0 \sum_j \cos(kj - \omega t), Z_j$. According to Fermi’s golden rule, the associated transition rate $\mathcal{I}(\omega, k)$ can be computed through the time-ordered correlation function as $\mathcal{I}(\omega, k)=-\frac{1}{\pi}\text{Im}~G(\omega,k)$~\cite{PhysRevA.95.043609,PhysRevA.81.063634}, where
  \begin{equation}
  \begin{aligned}
  G(\omega,k)=&-i \int dX dx dt~e^{-ikx+i\omega t}\\&\Big\langle \mathcal{T}\bar{\psi}\psi\Big(X+\frac{x}{2},\frac{t}{2}\Big) \bar{\psi}\psi\Big(X-\frac{x}{2},-\frac{t}{2}\Big)\Big\rangle,
  \end{aligned}
  \end{equation}
  with time-ordering operator $\mathcal{T}$. Similar to equal-time correlators, we analyze the OPE of $\bar{\psi}\psi(x_1,t_1)\bar{\psi}\psi(x_2,t_2)$, taking into account both spatial and temporal separations. The matrix element of this operator between dimer states receives contributions from three independent diagrams, the detailed computation of which are provided in the supplementary material \cite{SM}. Here, we present only the final result of the universal relation, valid in the regime $\omega/(u k^z) \sim O(1)$ and $\bar{n} \ll k \ll 1$:
  \begin{equation}
  \mathcal{I}(\omega, k)\approx \frac{1}{u^3k^{3z-1}}f(\omega/uk^z)\int dX~c(X),
  \end{equation}
  where $f(y)$ is a universal scaling function that depends only on $z$ (see Fig.~\ref{fig:DSF} for its behavior at different $\alpha$). The function $f(y)$ vanishes for $y < 2^{1-z}$, which corresponds to the kinematic threshold for creating a magnon pair with total momentum $k$. In particular, this excludes the divergence of the dynamical structure factor at $\omega/(u k^z) = 1$, associated with the excitation of a single magnon for $z \in (1/2,1)$, a feature characteristic of non-relativistic particles \cite{PhysRevA.81.063634}.

  \emph{\color{blue}Discussions.--} In this work, we establish universal relations in long-range quantum spin chains that connect equal-time correlators with the contact density. These relations extend the concept of universality beyond traditional ultracold atomic gases to a new class of quantum many-body systems: long-range quantum spin models. Our theoretical predictions, derived from effective field theory and the operator product expansion, are validated through matrix product state simulations of moderate system sizes. Furthermore, we derive a universal relation for the dynamical structure factor. Our results are directly accessible in state-of-the-art trapped-ion experiments, underscoring their experimental relevance. These findings open new avenues for exploring the interplay between few-body correlations and universal many-body phenomena in quantum simulation platforms that go beyond traditional nonrelativistic systems.

  We conclude our work with a few remarks. First, recent studies show that universal three-magnon states emerge for $\alpha \in (1.52, 1.88)$ in long-range spin chains \cite{2025arXiv250220759S}. These states are expected to contribute subleading corrections to the universal relations, analogous to three-body effects in identical bosons in three dimensions \cite{PhysRevLett.106.153005,PhysRevA.83.063614,PhysRevLett.112.110402}. Second, deriving the contact density for specific setups—such as a Bose-Einstein condensate of dilute magnon gases—would be an interesting direction, as it allows the exploration of novel many-body effects with a general dynamical exponent $z$. We leave a detailed analysis of these problems to future work.

  \textit{Acknowledgement.} 
  We thank discussions with Yi-Neng Zhou, Langxuan Chen, and Zeyu Liu. This project is supported by the Shanghai Rising-Star Program under grant number 24QA2700300 (PZ), the NSFC under grant 12374477 (PZ), the Innovation Program for Quantum Science and Technology 2024ZD0300101 (PZ) and 2023ZD0300900 (LF), and the Shanghai Municipal Science and Technology Major Project grant 24DP2600100 (NS and LF).

  \bibliography{ref.bib}

\end{document}


\title{Supplementary material: Universal Relations in Long-range Quantum Spin Chains }

  \author{Ning Sun}
  \affiliation{Department of Physics, Fudan University, Shanghai, 200438, China}

  \author{Lei Feng}
  \thanks{leifeng@fudan.edu.cn}
  \affiliation{Department of Physics, Fudan University, Shanghai, 200438, China}
  \affiliation{State Key Laboratory of Surface Physics, Fudan University, Shanghai, 200438, China}
  \affiliation{Institute for Nanoelectronic devices and Quantum computing, Fudan University, Shanghai, 200438, China}
  \affiliation{Shanghai Key Laboratory of Metasurfaces for Light Manipulation, Shanghai, 200433, China}
  \affiliation{Hefei National Laboratory, Hefei 230088, China}

  \author{Pengfei Zhang}
  \thanks{PengfeiZhang.physics@gmail.com}
  \affiliation{Department of Physics, Fudan University, Shanghai, 200438, China}
  \affiliation{State Key Laboratory of Surface Physics, Fudan University, Shanghai, 200438, China}
  \affiliation{Hefei National Laboratory, Hefei 230088, China}

  \date{\today}

  \begin{abstract}
  In this supplementary material, we provide details of the diagrammatic calculations.
  \end{abstract}

  \maketitle

  \section{Diagrams for equal-time correlators }

  \begin{figure}[t]
    \centering
    \includegraphics[width=0.65\linewidth]{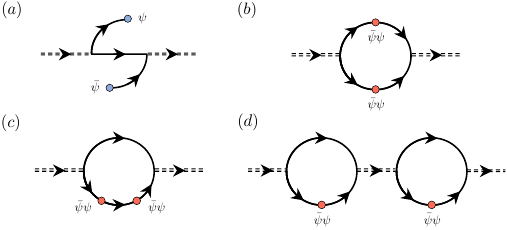}
    \caption{ We present Feynman diagrams for (a) the matrix element of $\bar{\psi}(x_1)\psi(x_2)$, (b) the matrix element of $\bar{\psi}\psi(x_1)\bar{\psi}\psi(x_2)$, (c-d) the calculation of the dynamical structure facgtor. In all diagrams, solid and dashed lines denote the bare propagators of $\psi$ and $d$, respectively.}
    \label{fig:sm}
  \end{figure}
  
  In this section, we compute the matrix element between dimer states to determine the OPE of equal-time correlators, considering an incoming dimer with energy $E$ and momentum $k$ and an outgoing dimer with energy $E'$ and momentum $k'$. We begin by examining the contribution to the $XX$-correlator. For clarity, we reproduce the diagram in FIG.~\ref{fig:sm}(a). Omitting the external lines, the diagram reads
  \begin{equation}
  \langle\text{out}|\bar{\psi}(x_1)\psi(x_2)|\text{in}\rangle=e^{-i \delta k X}(-i)^2\int \frac{dq_0}{2\pi}\frac{dq}{2\pi}\frac{i}{q_{0,+}-\epsilon_{q-\frac{\delta k}{2}}}\frac{i}{E_+-q_0-\epsilon_{\frac{k'+k}{2}-q}}\frac{i}{\delta E_++q_0-\epsilon_{q+\frac{\delta k}{2}}}e^{-iqx}.
  \end{equation}
  Here, we introduce $X=(x_1+x_2)/2$, $x=x_1-x_2$, $\delta k=k'-k$ and $\delta E=E'-E$. The factor of $(-i)$ comes from the vertex between one dimer and two magnons. We also define $E_+=E+i0^+$ as in the main text. The integration over $q_0$ leads to 
  \begin{equation}
  \langle\text{out}|\bar{\psi}(x_1)\psi(x_2)|\text{in}\rangle=e^{-i \delta k X}\int \frac{dq}{2\pi}\frac{1}{E_+-\epsilon_{\frac{k'+k}{2}-q}-\epsilon_{q-\frac{\delta k}{2}}}\frac{1}{E'_+-\epsilon_{\frac{k'+k}{2}-q}-\epsilon_{q+\frac{\delta k}{2}}}e^{-iqx}.
  \end{equation}
  We aim to extract the dominant short-range behavior, defined as $$1/\Lambda\ll|x|\ll \text{min}\{k^{-1},k'^{-1},(E/u)^{-1/z},(E'/u)^{-1/z}\}.$$ This corresponds to analyzing the integrand in the large-$q$ limit, which leads to
  \begin{equation}
  \langle\text{out}|\bar{\psi}(x_1)\psi(x_2)|\text{in}\rangle\approx e^{-i \delta k X}\int \frac{dq}{2\pi}\frac{1}{4\epsilon_{q}^2}e^{-iqx}=\frac{\Gamma(1-2z)\sin({\pi z})}{4u^2\pi}\frac{e^{-i \delta k X}}{|x|^{1-2z}}
  \end{equation}
  This is the result reported in the main text. Subleading corrections can also be extracted, which arises from a finite $E/u|q|^z$. This leads to 
  \begin{equation}
  (\text{Subleading})\approx e^{-i \delta k X}\int_{|q|\geq \epsilon } \frac{dq}{2\pi}\frac{E+E'}{8\epsilon_{q}^3}e^{-iqx}\approx e^{-i \delta k X}(E+E')\frac{\Gamma(1-3z)\sin({3\pi z/2})}{8u^3\pi}\frac{1}{|x|^{1-3z}}.
  \end{equation}
  Here, we drop a term that is $|x|$ independent by introducing a cutoff $\epsilon \sim \text{min}\{(E/u)^{1/z},(E'/u)^{1/z}\}$ near $|q|=0$. The leading term exhibits explicit dependence on $E+E'$, which can be matched with the contribution from the local operator $i\bar{d}\partial_t d - \partial_t \bar{d} d$. 

  Next, we consider the diagram that dominates the $ZZ$-correlator, as shown in FIG.~\ref{fig:sm}(b). Under the same setup, we obtain the matrix element 
  \begin{equation}
  \langle\text{out}|\bar{\psi}\psi(x_1)\bar{\psi}\psi(x_2)|\text{in}\rangle=e^{-i \delta k X}(-i)^2 \int \frac{dq_0}{2\pi}\frac{dq}{2\pi} \frac{dq_0'}{2\pi}\frac{dq'}{2\pi}\frac{1}{q_{0,+}-\epsilon_q}\frac{1}{E_+-q_{0}-\epsilon_{k-q}}\frac{1}{q_{0,+}'-\epsilon_{q'}}\frac{e^{i(q-q'+k'-k)x}}{E'_+-q_{0}'-\epsilon_{k'-q'}}.
  \end{equation}
  After integrating out $q_0$ and $q_0'$ and keeping the integrand to the leading order in $1/q$ and $1/q'$, we find  
  \begin{equation}
  \langle\text{out}|\bar{\psi}\psi(x_1)\bar{\psi}\psi(x_2)|\text{in}\rangle\approx e^{-i \delta k X} \left[\int \frac{dq}{2\pi} \frac{1}{2\epsilon_q}e^{iq x}\right]^2=\frac{\Gamma(1-z)^2\sin^2(\frac{\pi z}{2})}{4u^2\pi^2}\frac{e^{i (k-k')X}}{|x|^{2-2z}}.
  \end{equation}
  Similar to the analysis of the $XX$-correlator, the subleading contribution from a finite $E/(u|q|^z)$ yields a correction that scales as $1/|x|^{2-3z}$. Finally, we note that although diagrams (c) and (d) in Fig.~\ref{fig:sm} are also present, they do not contribute to non-trivial short-distance behavior in the calculation of the equal-time $ZZ$-correlator. We will return to this point later.

  \section{Diagrams for the dynamical structure factor }
  In this section, we compute the dynamical structure factor $\mathcal{I}(\omega, k)$ with $\omega,k>0$, which is related to the time-ordered correlation function as $\mathcal{I}(\omega, k)=-\frac{1}{\pi}\text{Im}~G(\omega,k)$, where 
  \begin{equation}
  \begin{aligned}
  G(\omega,k)=&-i \int dX dx dt~e^{-ikx+i\omega t}\Big\langle \mathcal{T}\bar{\psi}\psi\Big(X+\frac{x}{2},\frac{t}{2}\Big) \bar{\psi}\psi\Big(X-\frac{x}{2},-\frac{t}{2}\Big)\Big\rangle.
  \end{aligned}
  \end{equation}
  Unlike the calculation of the static correlation function, it is more convenient to work in the frequency–momentum domain and expand in large $\omega$ and $k$. The correlation function between dimer states receives three distinct contributions, corresponding to diagrams (b–d) in Fig.~\ref{fig:sm}.Motivated by the previous discussion, we focus on the leading-order contribution and set the energy and momentum of both the incoming and outgoing dimer to zero. The diagram (b) gives
  \begin{equation}
  G(\omega,k)^{(b)}_\text{dim}=(-i)^3L\int \frac{dq_0}{2\pi}\frac{dq}{2\pi}\frac{i}{q_{0,+}-\epsilon_q}\frac{i}{\omega_++q_0-\epsilon_{q+k}}\frac{i}{-q_{0,-}-\epsilon_q}\frac{i}{-\omega_--q_0-\epsilon_{q+k}}
  \end{equation}
  Here, we introduce the system size $L$ and $\omega_-=\omega-i0^+$. After performing the integration over $q_0$, we find
  \begin{equation}
  \begin{aligned}
  G(\omega,k)^{(b)}_\text{dim}&=L\int \frac{dq}{2\pi} \Bigg[ \frac{1}{\omega+u|q|^z-u|q+k|^z}\frac{1}{-2u|q|^z}\frac{1}{-\omega_--u|q|^z-u|q+k|^z}\\
  &-\frac{1}{\omega+u|q|^z-u|q+k|^z}\frac{1}{-2u|q+k|^z}\frac{1}{\omega_+-u|q|^z-u|q+k|^z}\Bigg].
  \end{aligned}
  \end{equation}
  The imaginary-part only arises from the pole at $\omega-u|q|^z-u|q+k|^z$. Introducing $\tilde{\omega}=\omega/u|k^z|$, the result is given by
  \begin{equation}
  \mathcal{I}(\omega, k)^{(b)}_\text{dim}=\frac{k^{1-3z}}{u^3}L \int \frac{dx}{2\pi}\frac{1}{4|x|^z|1+x|^z}\delta(\tilde{\omega}-|1+x|^z-x^z)\equiv\frac{k^{1-3z}}{u^3} L f^{(b)}(\tilde{\omega}).
  \end{equation}
  The result is non-zero only for $\tilde{\omega}> 2^{1-z}$, which corresponds to the kinematic threshold for creating a magnon pair with total momentum $k$. By matching to the matrix element of the contact operator, this result should be interpreted as the OPE relation $\mathcal{I}(\omega, k)\ni \frac{k^{1-3z}}{u^3} f^{(b)}(\tilde{\omega})\int dX c(X)$. Next, we compute the diagram (c), which gives
  \begin{equation}
  \begin{aligned}
  G(\omega,k)^{(c)}_\text{dim}&=(-i)^3L\int \frac{dq_0}{2\pi}\frac{dq}{2\pi}\frac{i}{q_{0,+}-\epsilon_q}\left[\frac{i}{-q_{0,-}-\epsilon_q}\right]^2\frac{i}{\omega_+-q_{0}-\epsilon_{q+k}}+\left(\omega \rightarrow -\omega\right)\\
  &=L\int \frac{dq}{2\pi} \frac{1}{(2u|q|^z)^2}\frac{1}{\omega_+-u|q|^z-u|q+k|^z}+\left(\omega \rightarrow -\omega\right).
  \end{aligned}
  \end{equation}
  Again, the imaginary-part arises from the pole at $\omega-u|q|^z-u|q+k|^z$. We find
  \begin{equation}
  \mathcal{I}(\omega, k)^{(c)}_\text{dim}=\frac{k^{1-3z}}{u^3} L\int \frac{dx}{2\pi}\frac{1}{4|x|^{2z}}\delta(\tilde{\omega}-|1+x|^z-x^z)\equiv\frac{k^{1-3z}}{u^3} L f^{(c)}(\tilde{\omega}).
  \end{equation}
  The last diagram in FIG. \ref{fig:sm}(d) reads 
  \begin{equation}
  \begin{aligned}
  G(\omega,k)^{(d)}_\text{dim}&=(-i)^5\times iT(\omega,k)L\left[\int \frac{dq_0}{2\pi}\frac{dq}{2\pi}\frac{i}{q_{0,+}-\epsilon_q}\frac{i}{-q_{0,-}-\epsilon_q}\frac{i}{\omega_+-q_{0}-\epsilon_{q+k}}\right]^2+\left(\omega \rightarrow -\omega\right)\\
  &=LT(\omega,k)\left[\int \frac{dq}{2\pi}\frac{1}{2u|q|^z}\frac{1}{\omega_+-u|q|^z-u|q+k|^z}\right]^2.
  \end{aligned}
  \end{equation}
  Here, $T(\omega,k)$ is the $T$-matrix for two-magnon scattering. We can parametrize it as 
  \begin{equation}
  T(\omega,k)^{-1}=-\frac{1}{2}\int \frac{dq}{2\pi}\left[\frac{1}{\omega_+-\epsilon_{\frac k2+q}-\epsilon_{\frac k2-q}}+\frac{1}{2u|q|^z}\right]\equiv\frac{k^{1-z}}{u}t(\tilde{\omega})^{-1}
  \end{equation} 
  This leads to 
  \begin{equation}
\mathcal{I}(\omega, k)^{(d)}_\text{dim}=-\frac{k^{1-3z}}{u^3} \frac{L}{\pi}~\text{Im}\left\{t(\tilde{\omega})\left[\int \frac{dx}{2\pi}\frac{1}{2|x|^{z}}\frac{1}{\tilde{\omega}_+-|x|^z-|x+1|^z}\right]^2\right\}\equiv\frac{k^{1-3z}}{u^3} L f^{(d)}(\tilde{\omega}) .
  \end{equation}
  Summing up contributions from all diagrams, we have 
  \begin{equation}
  \mathcal{I}(\omega, k)\approx \frac{k^{1-3z}}{u^3}\left[f^{(b)}(\tilde{\omega})+f^{(c)}(\tilde{\omega})+f^{(d)}(\tilde{\omega})\right] \int dX c(X)=\frac{k^{1-3z}}{u^3}f(\omega/uk^z)\int dXc(X).
  \end{equation}
  A numerical plot of the scaling function $f(\omega/uk^z)$ is provided in the main text.

  Finally, we note that diagrams (c) and (d) do not contribute to the $ZZ$-correlator. For both diagrams, after analytically continuing the frequency $\omega$ to the complex plane, each part of the expression is analytic in one half-plane. Consequently, for equal-time correlators, the additional integration over $\omega$ vanishes upon performing the contour integral.